# Timing and X-ray Spectral Features of Swift J1626.6−5156

Burçin İçdem[1], Sıtkı Çağdaş İnam[2], Altan Baykal[1]

burcinm@astroa.physics.metu.edu.tr; inam@baskent.edu.tr

## ABSTRACT

In this paper, we extend timing analysis of Baykal et al. (2010) of Swift J1626.6−5156 using RXTE-PCA observations between MJD 53724 and MJD 55113 together with a Chandra-ACIS dataset on MJD 54897 with a 20 ks exposure. We also present X-ray spectral analysis of these RXTE and Chandra observations. We find that the spin-up rate of the source is correlated with the X-ray flux. Using this correlation, we estimate the distance and surface magnetic field of the source as $\sim 15$kpc and $\sim 9 \times 10^{11}$Gauss respectively. From the spectral analysis, we found that power law index increases and Hydrogen column density decreases with decreasing flux.

## 1. Introduction

Swift J1626.6−5156 is a transient accretion powered pulsar with a spin period of ∼15 s (Palmer et al. 2005; Markwardt & Swank, 2005). It was first detected on 2005 December 18 with the Swift Burst Alert Telescope (BAT) (Palmer et al. 2005). Shortly after the 2005 outburst of the source, Reig et al. (2008) observed a ∼ 450 s X-ray flare, during which the pulsed fraction increased up to ∼ 70%. After the flare, the average count rate and the pulsed fraction were restored to their pre-flare values quickly. Reig et al. (2008) also reported a weak QPO feature with a characteristic frequency of 1 Hz and a fractional rms of 4.7%.

Proposed optical companion of the source (2MASS16263652-5156305, USNO-B1.0 0380-0649488) was found to show strong Hα emission, which is an indication of a Be star (Negueruela & Marco, 2006). As the infrared magnitudes of the companion is rather large for a Be star , i.e. the star is unusually faint in infrared band (J=13.5, H=13, K=12.6; Rea et al. 2006), Swift J1626.6−5156 is thought to be an unusual Be/X−ray binary system.

---

[1]METU, Physics Department, Ankara 06531 Turkey

[2]Başkent University, Department of Electrical and Electronics Engineering, Ankara 06530 Turkey

Baykal et al. (2010) studied long term monitoring RXTE (Rossi X-ray Timing Explorer) - PCA (Proportional Counter Array) observations of the source between MJD 53724 and 54410. They obtained timing solution of the source and found the orbital period to be 132.89 days. Baykal et al. (2010) also constructed long term pulse frequency history of Swift J1626.6−5156 and showed that the timescale of the X-ray modulations varied, which led to earlier suggestions (Reig et al. 2008; DeCesar et al. 2009) of orbital periods at about a third and half of the orbital period of Swift J1626.6-5156.

In this article, we extend timing analysis of Baykal et al. (2010) with the new analysis of RXTE-PCA data until MJD 55113 together with a Chandra-ACIS dataset with a 20 ks exposure. We also present X-ray spectral analysis of these RXTE and Chandra observations. In the next section, we describe the observations. In Sections 3 and 4, we present our timing and X-ray spectral analysis , and finally in Section 5, we discuss our results.

## 2. Observations

We analyzed data from Proportional Counter Array (PCA) onboard RXTE (Jahoda et al 1996) of Swift J1626.6−5156 between MJD 53724 and MJD 55113 with a total exposure of $\sim$ 449 ks, divided into 411 observations with exposures between $\sim$ 1 ks and $\sim$ 2 ks. We also note that this work includes a more detailed analysis of RXTE-PCA data between MJD 53724 and 54410 which were used before by Baykal et al. (2010).

The RXTE-PCA is an array of 5 Proportional Counter Units (PCU) sensitive to 2-60 keV energy range, with a total effective area of $\sim$ 7000 cm$^2$ and a field of view of $\sim$1° FWHM. During the analyzed RXTE-PCA observations, the number of active PCUs varied between 1 and 4. Before MJD 53964, there were 1-2 ks long observations every 2-3 days. After MJD 53964, observations were sampled in pairs, containing two consecutive 1-2 ks long observations separated by $\sim 0.3 - 0.6$ days. Each of these pairs were separated from each other by $\sim 9 - 10$ days. In timing analysis, we used all available layers of PCUs. For the spectral analysis, we used the layers of PCU2 only.

In addition to the RXTE-PCA data, we also used a Chandra AXAF CCD Imaging Spectrometer (ACIS; Garmire et al. 2003) observation of Swift J1626.6−5156 on MJD 54897 with an exposure of 20ks. This observation contains ACIS-S FAINT TE(timed exposure) mode data.

## 3. Timing Analysis

For timing analysis, we used 3-20 keV RXTE-PCA lightcurves with a time resolution of 0.375s which were generated from PCA GoodXenon data. Lightcurves were background subtracted using the background lightcurves generated with the background estimator models of RXTE team by using the standard PCA analysis tools (pcabackest in HEASOFT). Time columns in the lightcurves were also converted to those corresponding to the barycenter of the Solar system.

Pulse timing analysis between MJD 53724 and 54410 were already performed and corresponding pulse frequencies were obtained by Baykal et al. 2010. In this work, we extended this analysis which is based on the cross-correlation of pulse profiles represented harmonically (Deeter & Boynton 1985) to the data between MJD 54410 and 55113. We were not able to obtain pulse frequencies using the cross-correlation technique for the data after MJD $\sim$ 54750 since the pulse profiles obtained for these data were found to be statistically insignificant. In Figure 1, we present extended pulse frequency evolution of Swift J1626.6−5156. The pulse frequencies in Figure 1 were orbitally corrected using the orbital model of the source found by Baykal et al. (2010).

Using Chandra-ACIS observations, we also obtained a 20ks long lightcurve of the source with a time resolution of 0.44s. Crosscorrelating 8 pulse profiles obtained from $\sim$ 2.5 ks long segments of the Chandra-ACIS observations with a template pulse obtained from the whole observation, we obtained the spin period of the Swift J1626.6−5156 to be $(6.52059 \pm 0.00075) \times 10^{-2}$ Hz. We obtained 0.3-8 keV pulse profile of the source by folding of the whole Chandra-ACIS lightcurve with this frequency (see Figure 2).

By linear fitting of consecutive pulse frequencies with a time span of $\sim 10 - 140$ days, we obtained pulse frequency derivatives of the source between MJD 53724 and 54750. From the X-ray spectral analysis (see next section), we obtained unabsorbed 3-20 keV X-ray flux values corresponding to the pulse frequency derivative measurements. We found a correlation between X-ray flux and pulse frequency derivative (see Figure 3).

## 4. Spectral Analysis

In order to calculate 3-20 keV unabsorbed X-ray flux values of the source, we obtained X-ray spectra from the RXTE PCA observations corresponding to the spin frequency derivative measurements. We used the Standard-2 mode data, providing 128 channel spectra at 16 sec time resolution. Spectrum, background and response matrix files were created using FTOOLS 6.9 data analysis software. Energy channels corresponding to the 3-20 keV energy range were used to fit the spectra. We ignored photon energies lower than 3 keV and higher than 20 keV and 1%-2% systematic error was added to the errors (see Wilms et al. 1999; Coburn et al. 2000). Using Chandra-ACIS data, we also obtained 0.3-8 keV X-ray spectrum of the source (see Table 1).

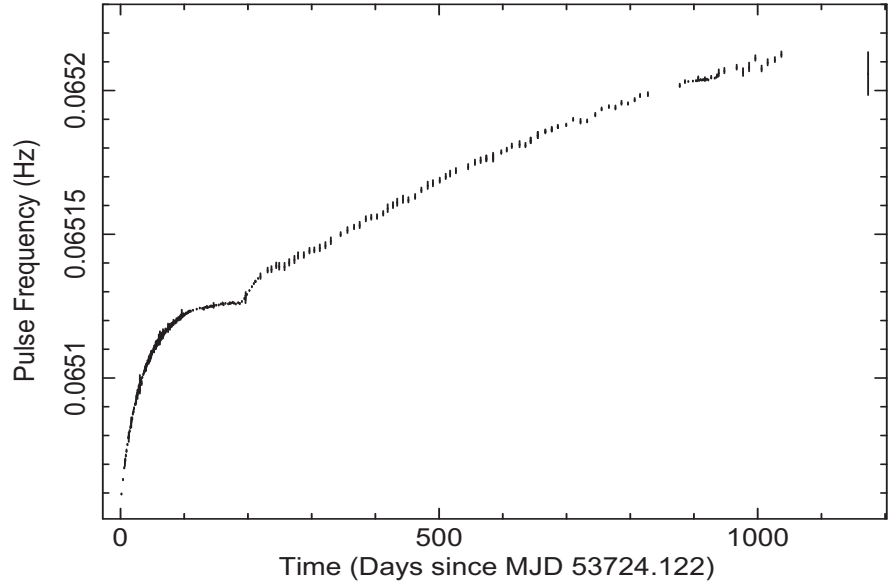

Fig. 1.— Pulse frequency evolution of Swift J1626.6−5156 after correcting for the binary orbital motion. The rightmost point corresponds to the Chandra-ACIS observation.

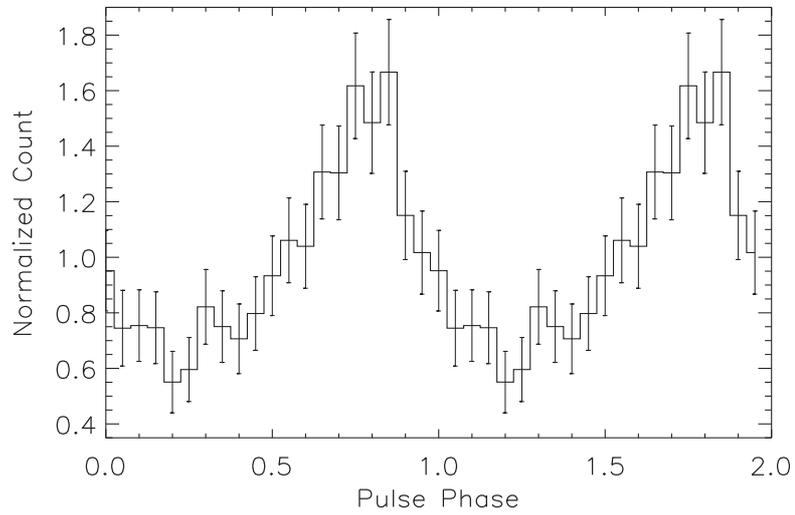

Fig. 2.— 0.3-8 keV pulse profile obtained from the Chandra ACIS observation.

We used an absorbed power law model with an high energy cutoff and an Iron line complex component modeled as a Gaussian peaking at $\sim 6.5$ keV. We could not resolve Cyclotron lines of the source around $\sim 10-20$keV as suggested by Coburn et al. (2006). Temporal variations of the unabsorbed X-ray flux, power law index and Hydrogen column density were presented in Figure 4. Average values of cut-off energy and e-fold energy values were found to be $\sim 15$keV and $\sim 10$keV respectively. To demonstrate long term spectral evolution of the source, we also constructed 3 long term RXTE-PCA spectra of the source (see Table 1).

## 5. Discussion

In Figure 3, we show that Swift J1626.6−5156 is an accretion powered pulsar that exhibit spin-up rate and X-ray flux correlation. Likewise, there are accretion powered pulsars that exhibit correlation between spin-up rate and X-ray flux: EXO 2030+375 (Parmar, White & Stella 1989; Wilson et al. 2002), A 0535+26 (Finger, Wilson & Harmon 1996; Bildsten et al. 1997), 2S 1417-62 (Finger, Wilson & Chakrabarty 1996; Inam et al. 2004), GRO J1744-28 (Bildsten et al. 1997), GRO J1750-27 (Scott et al. 1997), 2S 1845-024 (Finger et al. 1999), XTE J1543+568 (in't Zand, Corbet & Marshall 2001), SAX J2103.5+4545 (Baykal et al. 2002, 2007) and XMMU J054134.7-682550 (Inam et al. 2009).

Correlation between spin-up rate and X-ray flux can be considered as a sign of an accretion disc around the neutron star. In case of accretion via accretion disc, the Keplerian rotation of the disk is disrupted by the magnetosphere at the inner edge of the disc. Inside this radius, plasma is forced to move along the magnetic field lines to the magnetic poles of the neutron star. In case effect of radiation pressure is negligible (i.e. source luminosity is much smaller than Eddington luminosity and/or polar cap radiation is not directly beamed onto the accretion disk), the value of inner disk radius ($r_0$) is primarily related to the balance between the magnetic pressure and the material pressure exerted by the accreted material. Thus the inner disk radius is expected to decrease with increasing mass accretion rate ($\dot{M}$) and neutron star's mass ($M$) and decreasing magnetic moment ($\mu \simeq BR^3$, where $B$ is the surface magnetic field strength and $R$ is the neutron star radius). These dependences may approximately be expressed as (Pringle & Rees 1972; Lamb, Pethick, & Pines 1973)

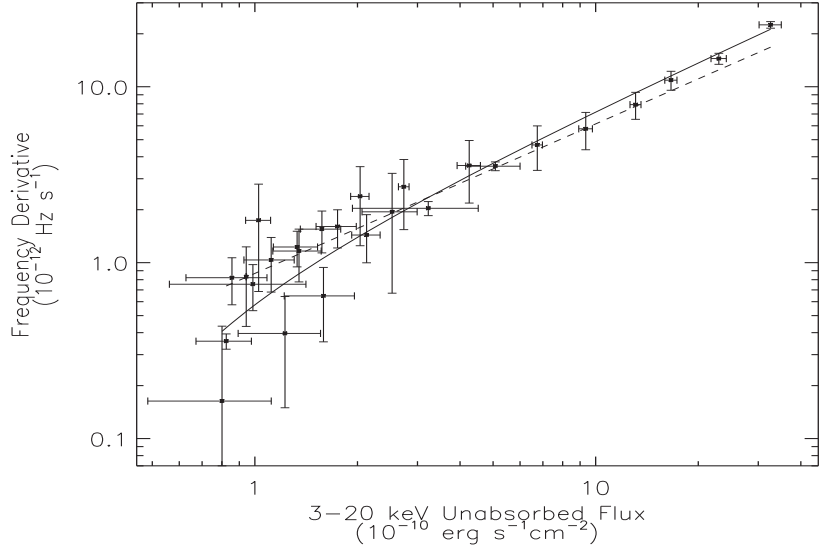

Fig. 3.— Frequency derivative of Swift J1626.6−5156 as a function of 3-20 keV unabsorbed X-ray flux obtained from RXTE-PCA observations. Solid and dashed lines correspond to the torque models with and without dimensionless torque parameter respectively.

Table 1: X-ray Spectral Parameters of Swift J1626.6−5156

| Instrument | RXTE-PCA | RXTE-PCA | RXTE-PCA | Chandra-ACIS |
|---|---|---|---|---|
| Time Interval (MJD-MJD) | 53724-53901 | 53901-54751 | 54751-55113 | 54897.31 |
| Exposure (ks) | 227 | 175 | 47 | 20 |
| $n_H$ ($10^{22}$cm$^{-2}$) | $6.31 \pm 1.31$ | $1.21 \pm 0.76$ | 0 (fixed) | $0.83 \pm 0.11$ |
| Power Law Index | $1.90 \pm 0.09$ | $2.00 \pm 0.05$ | $2.43 \pm 0.07$ | $0.94 \pm 0.12$ |
| Power. Law Norm. ($10^{-2}$photons.keV$^{-1}$.cm$^2$.s$^{-1}$) | $37.4 \pm 8.6$ | $6.43 \pm 0.82$ | $0.93 \pm 0.10$ | $(6.93 \pm 2.26) \times 10^{-3}$ |
| Iron Line Peak (keV) | $6.32 \pm 0.39$ | $6.57 \pm 0.11$ | $6.58 \pm 0.09$ | 6.58 (fixed) |
| Iron Line Sigma (keV) | $1.40 \pm 0.23$ | $0.76 \pm 0.12$ | 0 (fixed) | 0 (fixed) |
| Iron Line Norm. | $66.8 \pm 32.0$ | $7.41 \pm 2.41$ | $0.69 \pm 0.19$ | $(2.08 \pm 0.30) \times 10^{-2}$ |
| Cut-off Energy (keV) | $12.4 \pm 0.3$ | $15.4 \pm 0.5$ | $16.1 \pm 13.8$ | - |
| E-fold Energy (keV) | $21.7 \pm 2.4$ | $10.1 \pm 2.1$ | $8.57 \pm 8.50$ | - |
| Energy Range(keV-keV) | 3-20 | 3-20 | 3-20 | 0.3-8 |
| Unabsorbed X-ray Flux ($10^{-10}$.ergs.s$^{-1}$.cm$^{-2}$) | $14.1 \pm 0.1$ | $1.97 \pm 0.02$ | $0.13 \pm 0.13$ | $(9.43 \pm 1.26) \times 10^{-3}$ |
| Reduced $\chi^2$ / d.o.f | 0.51 / 32 | 1.05 / 32 | 0.65 / 34 | 1.25 / 34 |
| Systematic Error | 1% | 2% | - | - |

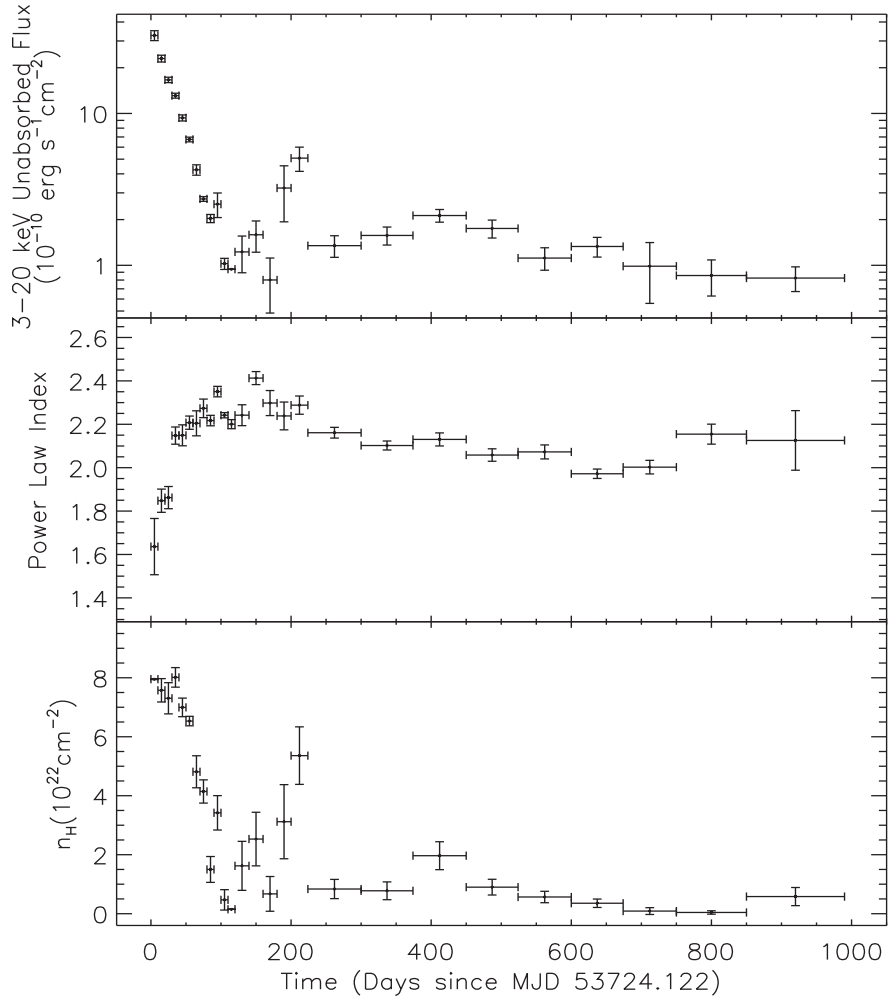

Fig. 4.— Temporal variations of 3-20 keV unabsorbed X-ray flux, power law index and Hydrogen column density.

$$r_0 = K\mu^{4/7} (GM)^{-1/7} \dot{M}^{-2/7}, \tag{1}$$

where $G$ is the gravitational constant and $K$ is a parameter which is of the order of unity. If $K$ equals 0.91, $r_0$ corresponds to Alfvén Radius for spherical accretion. The net torque exerted on the neutron star is estimated as (Ghosh & Lamb, 1979),

$$2\pi I \dot{\nu} = n(\omega_s)\dot{M}l_K \tag{2}$$

where $I$ is the moment of inertia of the neutron star, $l_K = (GMr_0)^{\frac{1}{2}}$ is the specific angular momentum added by the Keplerian disc to the neutron star at $r_0$;

$$n(\omega_s) \simeq 1.4(1 - \omega_s/\omega_c)/(1 - \omega_s) \tag{3}$$

is the dimensionless torque parameter which is a measure of the total (magnetic and material) torque exerted on the neutron star as a function of fastness parameter,

$$\omega_s = \nu/\nu_K(r_0) = (r_0/r_{co})^{3/2} = 2\pi K^{3/2} P^{-1}(GM)^{-5/7}\mu^{6/7}\dot{M}^{-3/7}, \tag{4}$$

where $r_{co} = (GM/(2\pi\nu)^2)^{1/3}$ is the corotation radius at which the stellar angular velocity equals Keplerian angular velocity. In Equation 3, $\omega_c$ is the critical fastness parameter which has been estimated as $\sim 0.35$ (Ghosh & Lamb 1979; Wang 1987; Ghosh 1993; Li & Wickramasinghe 1998; Torkelsson 1998; Dai & Li 2006).

The accretion leads to production of X-ray emission at the neutron star surface with a luminosity,

$$L = \eta\frac{GM\dot{M}}{R}, \tag{5}$$

where $\eta \leq 1$ is the efficiency factor. From Equations 1, 2, 5, spin-up rate is estimated as

$$\dot{\nu} \propto n(\omega_s)L^{6/7} = n(\omega_s)(4\pi d^2 F)^{6/7} \tag{6}$$

where $d$ is the distance to the source and $F$ is the X-ray flux. If the dimensionless torque ($n(\omega_s)$) is set to unity, this is the case in which material torques dominate. If the dimensionless torque is not unity, i.e. it is calculated using Equation 3, then this corresponds to the case in which the effect of magnetic torques is not neglected. In Figure 3, we present fits with $n = 1$ and the case for which the dimensionless torque is not unity (solid lines). We found that the model including non-unity dimensionless torque gives a better fit with a reduced $\chi^2$ of 1.05 compared to the fit with the unity dimensionless torque giving a reduced $\chi^2$ of 6.14.

Using Equation 6 and Figure 3, we estimated distance to the source as $\simeq 15$kpc. This leads to a surface magnetic field estimate of $\simeq 9 \times 10^{12}$ Gauss which is typical for accretion powered pulsars (see Coburn et al. 2002). This magnetic field value is also consistent with the previously suggested cyclotron lines around $\sim 10 - 20$keV of the source (Coburn et al. 2006).

From Figure 2, it is shown that pulses from Swift J1626.6−5156 do not cease even after $\sim 1200$ days from the outburst. This indicates that the source still accretes matter without any significant accretion geometry change.

In Figure 4, we found that power law index is anticorrelated with the X-ray flux. This indicates that the spectrum becomes softer with decreasing X-ray flux which is expected and may be interpreted as a consequence of mass accretion rate changes without a need of an accretion geometry change (Meszaros et al. 1983; Harding et al. 1984). Similar anticorrelation was also found for 2S 1417-62 (Inam et al. 2004) and SAX J2103.5+4545 (Baykal et al. 2007). Figure 4 also reveals that Hydrogen column density is correlated with the X-ray flux. This correlation was also seen in 2S 1417-62 and was thought to be due to the fact that matter concentration around the neutron star should be related to the mass accretion rate in Be/X-ray pulsar systems (Inam et al. 2004).

We acknowledge research project TBAG 109T748 of the Scientific and Technological Research Council of Turkey (TÜBİTAK).